\documentclass[preprint,showkeys]{revtex4}
\usepackage{amsmath}
\usepackage{amssymb}
\begin{document}

\title{Non-Gaussian continuous-variable teleportation}  
\author{Paulina Marian}  
\author{ Tudor A. Marian}
\affiliation{ Centre for Advanced  Quantum Physics,
University of Bucharest, P.O.Box MG-11,
R-077125 Bucharest-M\u{a}gurele, Romania}

\begin{abstract}
We have recently shown that the output field in the Braunstein-Kimble protocol of teleportation is a superposition of two fields: the input one and a field created by Alice's measurement and by displacement of the state at Bob's station by using the classical information provided by Alice. We study here the noise added by teleportation and compare its influence in the Gaussian and non-Gaussian settings. 
\end{abstract}

\maketitle
\section{Introduction}	

In the Braunstein-Kimble protocol for teleporting one-mode states of the electromagnetic field \cite{BK}, two distant operators, Alice at a sending station, and Bob 
at a receiving terminal, share a two-mode entangled quantum state 
and exploit its  nonlocal character as a quantum resource for teleporting a single-mode state whose density operator is denoted by $\rho_{in}$. Mode $1$ of the shared resource having the density operator $\rho_{AB}$ is given to Alice and 
mode $2$ is given to Bob. First Alice  performs 
a complete projective measurement on the joint system described by the three-mode product state 
$\rho_{in}\otimes\rho_{AB}$
and then conveys its outcome to Bob via a classical 
communication channel. As a consequence of Alice's measurement, 
the total state of the three-mode system collapses. Finally, Bob makes use of the information 
transmitted classically by Alice to transform his state into 
an output that is a replica of the original unknown input. Unfortunately, a perfect replica is obtained only for an ideally entangled state  $\rho_{AB}$. In general, teleportation in the continuous-variable settings is a noise-generating process \cite{PT}.  A comprehensive account for the role of teleportation
 in the context of continuous-variable quantum information  is given in the reviews \cite{BL,F1}.
The present paper is a continuation of our work on Braunstein-Kimble protocol in the characteristic-function (CF) description \cite{PT}. We focus here on describing the distorsion of the teleported state in terms of the properties of the two-mode resource state. In Sec. 2 we review the steps of the Braustein-Kimble protocol and derive the factorized CF of the teleported state. We discuss in Sec.3 the properties of the distorting state by using the relation between its normally ordered correlation functions and two-mode correlation functions of the resource state. We find that the conclusions drawn in our paper \cite{PT} for a Gaussian resource are valid in general. In Sec.4 we show that the mean occupancy in the distorting field equals the EPR-uncertainty of the resource state. This gives us reasons for using it as a measure of teleportation accuracy.
Finally we discuss the case of pure resource states of given entanglement and show that the two-mode squeezed vacuum state (SVS) generates the minimal noise in the teleportation output.
\section{Teleportation revisited}

 Recall the one-to-one correspondence between
 the density operator $\rho$ of a $ n$-mode field state 
and its  CF defined as the expectation value of the $ n$-mode displacement operator 
$\chi(\lambda_1,\lambda_2\cdots \lambda_n):={\rm Tr}[\rho D(\lambda_1)
D(\lambda_2)\cdots D(\lambda_n)]$. Here $D(\alpha)=\exp{(\alpha \hat a^{\dag}-\alpha^* \hat a)}$ is a Weyl 
displacement operator and $\hat a$ denotes the annihilation operator. The density operators of the states involved in the protocol, $\rho_{in}$ and $\rho_{AB}$,  were written as Weyl expansions,
\begin{eqnarray}
\rho_{in}=\frac{1}{\pi}\int {\rm d}^2 \lambda \;\;\chi_{in} (\lambda)
D(-\lambda),\label{cf1}
\end{eqnarray}
and
\begin{eqnarray}
\rho_{AB}&=&\frac{1}{\pi^2}\int {\rm d}^2 \lambda_1 {\rm d}^2 \lambda_2\;\;
\chi_{AB}(\lambda_1,\lambda_2)D_1(-\lambda_1)D_2(-\lambda_2).\nonumber \\&&
\label{cf2}
\end{eqnarray}
Here $\chi_{in}(\lambda)$ is the CF of the one-mode input state and
$\chi_{AB}(\lambda_1,\lambda_2)$ describes the two-mode resource state. The steps of the Braunstein-Kimble protocol are summarized as follows. 
\begin{itemize}
\item quantum measurement of the variables 
\begin{equation} \hat Q_m=\hat q_{in}- \hat q_1,\;\;
 \hat P_m=\hat p_{in}+\hat p_1 \label{measure}\end{equation}
{\rm with the canonical operators}: $\hat q_j=(\hat a_j+\hat a_j^{\dag})/{\sqrt{2}},\;\;
 \hat p_j=(\hat a_j
-\hat a_j^{\dag})/(\sqrt{2}i)$. $\hat a_j $ and $\hat a_j^{\dag}$  
$(j=1,2)$ are the amplitude operators 
for the entangled modes of the state $\rho_{AB}$.
 The common eigenfunctions of the commuting observables $\hat Q_m,\hat P_m$ have the following expansion
 \begin{equation} |\Phi_{in,A}(q,p)\rangle=\frac{1}{(2 \pi)^{1/2}}
\int\limits_{-\infty}^{\infty}{\rm d}\eta{\rm e}^{ip\eta}
|q+\eta\rangle_{in}\otimes|\eta\rangle_A,\end{equation}
where the pair $(q,p)$ denotes the outcome of the measurement.
\item classical communication of the measurement's results (performed by Alice to Bob)
The state at Bob's side predicted by quantum mechanics after Alice's measurement  is
\begin{equation}
\rho_B(q,p)\sim 
{\rm Tr}_{in,A}\left\{\left[|\Phi(q,p)\rangle
\langle\Phi(q,p)|\otimes I_B\right]
(\rho_{in}\otimes\rho_{AB})\right\}.\end{equation}
 \item displacement of the state at Bob's location with the parameter $\mu=q+i p$
\end{itemize}
Following the protocol step by step  we have finally  reached the following factorization formula for the CF of the teleported state $\chi_{out}(\lambda)$ 
\begin{eqnarray}
&&\chi_{out}(\lambda)= \chi_{in}(\lambda)\chi_{AB}(\lambda^*,\lambda).
\label{telCF}
\end{eqnarray}
Therefore $\chi_{out}(\lambda)$ is the product between the
 CF of the input state $\chi_{in}(\lambda)$ and a function only depending on the 
properties of the two-mode resource state and the geometry of measurement \cite{PT}. 
 We have interpretated Eq.\ (\ref{telCF}) as describing the superposition between
the input field $in$ and a single-mode one reduced
from the entangled 
$AB$ field 
by the measurement performed by Alice followed by the 
phase-space translation 
 performed by Bob.
We have identified the function $\chi_{AB}(\lambda^*,\lambda)$
 as the normally ordered CF of a one-mode {\em distorting field} whose density operator is denoted by $\rho_M$:
\begin{equation}\chi^{(N)}_{M}(\lambda)= \chi_{AB}(\lambda^*,\lambda).\label{M}\end{equation}
 Equation \ (\ref{telCF}) is valid for arbitrary input and resource states. According to
 Eq.\ (\ref{M})  the properties of 
the distorting field state $\rho_M$ are fully determined by the  two-mode 
resource state. Thus, if the entangled  state ${\rho}_{AB}$ is a two-mode Gaussian state, 
then any single-mode Gaussian input is teleported as a single-mode 
Gaussian output.  In our paper \cite{PT} we focused on the Gaussian resource state case and derived  several properties of the  state $\rho_M$. We have written the covariance matrix (CM) of the state $\rho_M$ by making explicitly use of the Gaussian character of the resource state ${\rho}_{AB}$.  The input state distortion through  teleportation was described 
by the average photon number of the measurement-induced field $\rho_M$ that we called added noise. In the case of symmetric Gaussian resource states we have found a relation between the optimal added noise and the minimal EPR correlations used to define inseparability.
Our principal aim in the present paper is to deepen the relation between the properties of the two-mode resource state and those of the distorting field $M$.
\section{Second-order correlations}
According to the definition of a normally ordered CF \cite{Gl} we have for the
one-mode distorting state (amplitude operators $\hat a,\hat a^{\dag}$)
\begin{equation}
\chi^{(N)}_M(\lambda)=\sum_{l, m=0}^{\infty}
\frac{1}{l!m!}
\lambda^{l}(-\lambda^{*})^{m}\langle(\hat a^{\dag})^{l}\hat a^{m}\rangle_M, \label{2.3}
\end{equation}
 From the similar expansion of the two-mode CF of the state ${\rho}_{AB}$ we find
\begin{eqnarray}
\chi_{AB}(\lambda^*,\lambda)&=&\exp{(-|\lambda|^2)}\sum_{s,r,l,m}
\left(\begin{array}{c}l\\ r
\end{array}\right)\left(\begin{array}{c}m\\ s
\end{array}\right)\frac{(-1)^{r+s}\lambda^l (-\lambda^*)^m}{l! m!}
\nonumber\\  &&\times
\langle (\hat a_1^{\dag})^{s}\, (\hat a_2^{\dag})^{l-r}\,\hat a_1^{r}\,\hat  a_2^{m-s}\rangle_{AB}.
\label{3}\end{eqnarray}
By using Eq.\ (\ref{telCF}) we could find  relations between the correlation functions
of the distorting state and two-mode correlation functions of the resource state. Let us now suppose that the two-mode resource state is undisplaced. Our first
goal is to write the CM of the one-mode distorting field $\rho_M$ defined as
\begin{eqnarray}
{\cal V}_M=\left(\begin{array}{cc}\sigma(qq)&\sigma(qp)\\
\sigma(qp)&\sigma(pp)\end{array}\right).
\label{vm}\end{eqnarray} 
In Eq.\ (\ref{vm}) we have denoted by $\sigma(qq)=\langle \hat q^2\rangle_M,\;\sigma(pp)=\langle \hat p^2\rangle_M,\;\sigma(qp)=\langle \hat q \hat p\rangle_M$, the second-order correlations of the canonical operators
$\hat q=(\hat a+\hat a^{\dag})/{\sqrt{2}},\;\hat p=(\hat a-\hat a^{\dag})/(\sqrt{2}i)$,
where $\hat a $ and $\hat a^{\dag}$,   are the amplitude operators 
of the state $M$. 
Equation \ (\ref{telCF}) gives us via Eqs.\ (\ref{2.3}) and \ (\ref{3})
\begin{equation}\langle \hat a^2\rangle_M=\langle\hat  a_2^2\rangle_{AB}+\langle (\hat a_1^{\dag})^2\rangle_{AB}-2\langle \hat a_1^{\dag} \hat a_2\rangle_{AB},\label {11}\end{equation}
\begin{equation}\langle\hat  a^{\dag} \hat a\rangle_M=1+\langle \hat a_1^{\dag} \hat a_1\rangle_{AB}+\langle\hat  a_2^{\dag}\hat  a_2\rangle_{AB}-\langle \hat a_1 \hat a_2\rangle_{AB}-\langle \hat a_1^{\dag}\hat  a_2^{\dag}\rangle_{AB},\label {12}\end{equation}
 such that the entries of the CM of the distorting state $M$ are
\begin{eqnarray}\sigma(qq)&=&\frac{1}{2}+\sigma(q_2q_2)+
\sigma(q_1q_1)- 2 \sigma(q_1q_2).\label{a}\end{eqnarray}
\begin{eqnarray}\sigma(qp)
&=&\sigma(q_2p_2)-\sigma(q_1p_1)
+\sigma(q_2p_1)-
\sigma(q_1p_2).\label{b}\end{eqnarray}
\begin{eqnarray}\sigma(pp)&=&\frac{1}{2}+\sigma(p_2p_2)+
 \sigma(p_1p_1)+2 \sigma(p_1p_2).\label{c}
\end{eqnarray}
We have thus expressed the CM of the state $\rho_M$ in terms of the correlations of the canonical operators $\hat q_j,\; \hat p_j,
(j=1,2)$ of the two-mode resource state $\rho_{AB}.$
Equations \ (\ref{a})--\ (\ref{c}) are valid for arbitrary two-mode resource state (Gaussian and non-Gaussian).
For later convenience let us introduce two commuting operators  $\hat Q, \hat P$  closely related
 to those measured by Alice:
\begin{eqnarray} \hat Q (\hat q_1, \hat q_2):=\hat Q_m (\hat q_{1}\rightarrow \hat q_1, \hat q_{in}\rightarrow \hat q_2)=\hat q_2-\hat q_1\end{eqnarray}
\begin{eqnarray}\hat P(\hat p_1, \hat p_2)=\hat P_m(\hat p_1\rightarrow \hat p_1, \hat p_{in}\rightarrow \hat p_2)=\hat p_1+\hat p_2.\end{eqnarray}
The correlations  \ (\ref{a})--\ (\ref{c}) are written now
\begin{equation}\sigma(qq)=\frac{1}{2}+\langle \hat Q^2\rangle,\;\;\sigma(qp)=\langle 
\hat Q \hat P\rangle,\;\;\sigma(pp)=\frac{1}{2}+\langle 
\hat P^2\rangle.\label{QP}\end{equation}
Two conclusions arise from the new aspect of the entries of the CM \ (\ref{vm}):
\begin{enumerate}
\item The Robertson-Schr\"odinger uncertainty relation, 
\begin{eqnarray}\det{\cal V}_M=\sigma(qq)\sigma(pp)- 
(\sigma(qp))^2 \geq \frac{1}{4},\end{eqnarray}
is verified because
\begin{eqnarray}\langle 
\hat Q^2\rangle\langle \hat P^2\rangle-\langle 
\hat Q \hat P\rangle^2\geq 0.\label{UR1}\end{eqnarray}
\item The distorting state is not squeezed: ${\cal V}_M\geq \frac{1}{2}I_2. $
\end{enumerate}
 \section{Accuracy of teleportation. Added noise}
Originally, the quality of the teleportation protocol was quantified
 by the overlap of the $in$ and 
$out$ states for pure $in$ states \cite{BK} or the Uhlmann fidelity 
for mixed Gaussian states \cite{PTH03,Ban2}. So defined, 
the {\em fidelity of teleportation} depends on the input state:
$${\cal F}(in,out)=\frac{1}{\pi}\int {\rm d}^2 \lambda \chi_{in}^*(\lambda) \;\chi_{out}(\lambda).$$
In particular,
  the fidelity of teleportation for a coherent state is written via Eq.\ (\ref{telCF})
\begin{equation}{\cal F}_{coh}=\frac{1}{\pi}\int {\rm d}^2 \lambda \exp{(-|\lambda|^2)}
\chi_{AB}(\lambda^*,\lambda),\label{fc1}\end{equation}
or, equivalently
\begin{equation}{\cal F}_{coh}={\cal Q}_M (0).\label{fc2}\end{equation}
In Eq.\ (\ref{fc2}), ${\cal Q}_M (0)$ is the expectation value of the density operator of the state $M$ in vacuum, namely is the ${\cal Q}(\alpha)$-function \cite{Gl} of this state at  $\alpha=0$.

Following Refs.\cite{HK,F,MF} we evaluate the teleportation quality in terms 
of the mean occupancy in the remote field $\rho_M$ which can be seen as 
the amount of noise distorting the properties of the input field state. From Eq.\ (\ref{12}) via Eq.\ (\ref{QP}) we get
\begin{equation}\langle\hat  a^{\dag} \hat a\rangle_M
=\frac{1}{2}\left[\langle 
\hat Q^2\rangle+\langle \hat P^2\rangle\right].\label{EPR}\end{equation}
The r.h.s. of Eq.\ (\ref{EPR}) is in fact the EPR-uncertainty generally defined in the undisplaced two-mode case as
\begin{equation} \Delta_{EPR}(\rho):=\frac{1}{2}\left[\langle (\hat q_2-\hat q_1)^2+\langle (
\hat p_1+\hat p_2)^2 \rangle\right].\label{EPR1}\end{equation}
We can now formulate the main results of this paper.

{\em Theorem 1: The amount of noise distorting the properties of the input field state is rigourously equal to the EPR--uncertainty of the resource state 
$\rho_{AB}$.}
\begin{equation}\langle\hat  a^{\dag} \hat a\rangle_M=\Delta_{EPR}(\rho_{AB}).\label{EPR2}\end{equation}
Note that this is valid far an arbitrary two-mode resource state. When the condition $\Delta_{EPR}(\rho_{AB})<1$ is met the state $\rho_{AB}$ presents non-local correlations. Therefore the teleportation process generates less noise in the output state when the non-locality of the resource state expressed by the EPR-uncertainty 
\ (\ref{EPR1}) is stronger ($\Delta_{EPR}(\rho_{AB})$ is smaller). 

We discuss now the case of pure two-mode resource states.
 According to  Giedke {\em et al.} \cite{Giedke}, there is a direct relation between the amount of entanglement and the EPR-uncertainty of pure two-mode states. Thus, among all pure two-mode states (Gaussian and non-Gaussian), the squeezed vacuum state (SVS) has the minimal amount of entanglement at a
prescribed EPR--uncertainty. Otherwise said, among all pure states with the same entanglement the SVS has minimal
$\Delta_{EPR}$ (maximal EPR--correlations). Application of this  important result  leads us to a strong interpretation of 
Eq.\ (\ref{EPR2}):

 {\em Theorem 2: The minimal noise added in teleportation with pure two-mode resource states having the same entanglement is realized
by the SVS.}

To conclude, in this paper we have shown that the quality of the continuous-variable teleportation is determined by the amount of non-locality measured by the EPR-uncertainty of the resource state. As a consequence of the factorization formula \ (\ref{telCF}) this characterization of the efficiency of the teleportation process is valid for arbitrary input one-mode states. We could apply the strong theorem proved in Ref.\cite{Giedke} to show that SVS  generates the minimal noise in the teleportation output when comparing all pure resource states of given entanglement.

\section*{Acknowledgments}
This work was supported by the Romanian 
ANCS through Grant No.IDEI-995/2007 for the University of Bucharest.

\end{document}